\documentclass[10pt,conference]{IEEEtran}
\usepackage{amsmath,amssymb,eucal,graphicx}
\usepackage{epsfig}
\usepackage{exscale}

\def\LB{\left(}         %Left bracket
\def\RB{\right)}        %Right bracket

\def\tfb{T_{\rm fb}}
\def\nt{N_t}

\newfont{\bbb}{msbm10 scaled 500}

\newfont{\bb}{msbm10 scaled 1100}
\newcommand{\CC}{\mbox{\bb C}}

\newcommand{\EE}{\mbox{\bb E}}

\newcommand{\hv}{{\bf h}}

\newcommand{\sv}{{\bf s}}

\newcommand{\vv}{{\bf v}}
\newcommand{\xv}{{\bf x}}

\newcommand{\zv}{{\bf z}}

\newcommand{\Id}{{\bf I}}

\newcommand{\Cc}{{\cal C}}

\newcommand{\Kc}{{\cal K}}
\newcommand{\Lc}{{\cal L}}

\newcommand{\Nc}{{\cal N}}

\newcommand{\SNR}{{\sf SNR}}

\DeclareFontFamily{U}{cmfi}{}
\DeclareFontShape{U}{cmfi}{m}{n}{ <-> cmfi10 }{}
\DeclareSymbolFont{CMFI}{U}{cmfi}{m}{n}

\begin{document}

\vspace{-3cm}
\title{Optimized Training and Feedback for MIMO Downlink Channels}

\author{\authorblockN{Mari Kobayashi}
\authorblockA{SUPELEC \\
Gif-sur-Yvette, France\\
%Email: {\tt mari.kobayashi@supelec.fr}\\
}
\and
\authorblockN{Nihar Jindal}
\authorblockA{University of Minnesota \\
Minneapolis MN, 55455 USA
}
\and
\authorblockN{Giuseppe Caire}
\authorblockA{University of Southern California\\
Los Angeles CA, 90089 USA\\
%Email: {\tt caire@usc.edu}\\
}}
%Email: {\tt nihar@umn.edu}\\}
%\and
%\authorblockN{Niranjay Ravindran,}
%\authorblockA{University of Minnesota \\
%Minneapolis MN, 55455 USA\\
%Email: {\tt ravi0022@umn.edu}\\
%}}

\maketitle
\begin{abstract}
We consider a MIMO fading broadcast channel where channel state
information is acquired at user terminals via downlink training and
channel feedback is used to provide transmitter channel
state information (CSIT) to the base station. The feedback channel
(the corresponding uplink) is modeled as an AWGN channel, orthogonal across
users.  The total bandwidth consumed is the sum of the
bandwidth/resources used for downlink training, channel feedback, and data transmission.
Assuming that the channel follows a block fading model and that
zero-forcing beamforming is used, we optimize the net achievable rate
for unquantized (analog) and quantized (digital) channel feedback.
The optimal number of downlink training pilots is seen to be essentially
the same for both feedback techniques, but digital feedback is shown
to provide a larger net rate than analog feedback.

\end{abstract}

%%%%%%%%%%%%%%%%%%%%%%%%%%%%%%%%%%%%%%%%%%%%%%%%%%%%%%%%%%
\section{Introduction}

We consider a MIMO Gaussian broadcast channel modeling the downlink of a system where a Base Station (BS)
has $\nt$ antennas and serves $K$ single-antenna User Terminals (UTs).
A channel use of such system is described by
\begin{equation} \label{model}
y_k = \hv_k^H \xv + z_k, \;\; k = 1,\ldots,K
\end{equation}
where $y_k$ is the channel output at UT $k$,
$z_k \sim \Cc\Nc(0,N_0)$ is the corresponding
Additive White Gaussian Noise (AWGN),
$\hv_k \in \CC^\nt$ is the vector of channel coefficients from the BS antenna array to the
$k$-th UT antenna and $\xv$ is the vector of channel input symbols transmitted
by the BS, subject to the average power constraint $\EE[|\xv|^2] \leq P$.
We use $\rho$ to denote the nominal SNR: $\rho \triangleq \frac{P}{N_0}$.
We assume a block fading model, i.e., the channel remains constant over a coherence interval
of $T$ channel uses.  Albeit suboptimal, we focus on zero-forcing (ZF) beamforming with $K=\nt$ users for its analytical
tractability.

In order to perform zero-forcing beamforming (or any other multi-user MIMO strategy),
the BS must have an accurate estimate of the channel to each UT.
Such information is generally acquired in a two-step process: each UT
first estimates its own downlink channel during a common downlink training phase, after
which each UT transmits its channel estimate over a feedback channel (on the
corresponding uplink) to the BS.  The rates achievable with
zero-forcing beamforming depend critically on the quality of the CSI available to the BS,
but high quality CSI can only be achieved if (a) each UT is able to accurately estimate
its own channel (i.e., by using a suitably long downlink training phase), and (b) the
process of channel feedback does not add too much additional distortion to the channel
information.

In this work, we attempt to determine the optimum fraction of resources that should
be dedicated to training and feedback with the criterion being the \textit{net}
spectral efficiency achievable on the downlink. Training consumes downlink bandwidth while
channel feedback consumes uplink bandwidth. From a high-level system perspective,
increasing either training or feedback effectively takes away from the bandwidth
available for actual data transmission. The net spectral efficiency is therefore
the transmission rate reflecting the overhead due to the downlink bandwidth consumed for training and the
uplink bandwidth used for channel feedback.

We utilize our earlier work in which the downlink spectral efficiency was tightly bounded as a function of the amount of training and
feedback \cite{Submitted}.  This allows the net spectral efficiency per UT, accounting for training/feedback resources, to be accurately lower bounded as
\begin{equation}\label{Corollary3}
    R_k \geq \left(1-\frac{T_1 + \tfb}{T}\right) \left( R^{\rm ZF} - \overline{\Delta R} \right)
\end{equation}
where $T_1$ and $\tfb$ are the number of channel symbols used for training and feedback,
respectively, per coherence block of length $T$.\footnote{Using $T_1$ and $\tfb$ symbols
per length $T$ block is fully equivalent to using a fraction $\frac{T_1}{T}$ and $\frac{\tfb}{T}$
of the total bandwidth for training and feedback.}  The quantity $R^{\rm ZF}$ denotes the rate
achievable with ideal CSI while $\overline{\Delta R}$ is the rate gap due to imperfect CSI.
This rate gap depends on the particular feedback strategy used, as specified in \cite{Submitted},
but is decreasing in both $T_1$ and $\tfb$.  The objective of this paper is to maximize
the net spectral efficiency with respect to $T_1$ and $\tfb$ for the cases of analog and digital
feedback, and to understand how the optimal values of $T_1$ and $\tfb$ as well as the
optimized net spectral efficiency depend on various system parameters of interest (e.g., blocklength $T$,
signal-to-noise ratio, and $\nt$).

The present work is an extension of \cite{training_isit08}, where the
same optimization was investigated for the case of analog (unquantized) feedback
over a shared MIMO MAC feedback channel.  On the other hand, here we consider
both analog and digital feedback techniques and focus primarily on an AWGN model
for the feedback channel.  Interested readers can refer to  \cite{training_isit08}
for a discussion of other prior work on this general topic.

%%%%%%%%%%%%%%%%%%%%%%%%%%%%%%%%%%%%%%%%%%%%%%%%%%%%%%%%%%
\section{Channel State Estimation and Feedback}\label{sect:TrainingFB}

In this section we describe the basic training and channel feedback scheme
that allows CSI to be acquired.

1) Common downlink training: $T_1$ shared pilot symbols (essentially
$\frac{T_1}{\nt}$ pilots per BS antenna) are transmitted to allow all UTs to estimate their
downlink channel vectors $\{\hv_k\}$ based on the observation
\begin{equation} \label{training-phase-1-rx}
\sv_k = \sqrt{\frac{T_1P}{\nt}}\ \hv_k + \zv_k
\end{equation}
where $\zv_k \sim \Cc\Nc(0,N_0 \Id)$.
Each UT performs linear MMSE of $\hv_k$ from the observation $\sv_k$, which
results in a per-coefficient estimation error with variance \cite[Equation 7]{Submitted}
\begin{eqnarray}
\frac{1}{1 + \left(\frac{T_1}{\nt}\right) \rho }.
\end{eqnarray}

2) Channel feedback: Each UT feeds back its channel estimation immediately after
the training phase.  We focus on the scenario where the feedback channel is modeled
as an AWGN channel with the same signal-to-noise ratio $\rho$, identical to the nominal
downlink SNR.  Because UT's are assumed to access the feedback channel orthogonally, a total
of $\tfb$ channel symbols translates into $\frac{\tfb}{\nt}$ feedback channel uses
per UT.  The different feedback strategies are described in Section \ref{sect:Optimization}.

From the feedback received from each of the UT', the BS obtains the channel estimate $\widehat{\hv}_1,\ldots, \widehat{\hv}_{\nt}$.
The imperfection in the CSI available to the BS stems from two sources: the channel estimation error during the common
training phase, and the distortion incurred during the feedback phase.
For analog feedback the distortion is due to additive noise in the feedback channel, while
for digital feedback it consists of the quantization error as well
as possible errors while transmitting bits over the feedback channel.

If the beamforming vectors $\widehat{\vv}_1,\ldots, \widehat{\vv}_{\nt}$ are selected by
using zero-forcing on the basis of the imperfect channel estimates
$\widehat{\hv}_1,\ldots, \widehat{\hv}_{\nt}$, the following per-UT
rate is achievable if equal-power (across UT's) Gaussian inputs are used:
\begin{eqnarray} \label{eq-rate}
\mathbb{E} \left[ \log \left(1 + \frac{ |\hv_k^H \hat{\vv}_k|^2 \frac{\rho}{\nt} }
{1 + \frac{\rho}{\nt} \sum_{j \ne k} |\hv_k^H \hat{\vv}_j|^2 } \right) \right],
\end{eqnarray}
assuming each UT is aware of its received SINR.\footnote{Such knowledge can be
acquired through an additional dedicated training round as discussed in \cite{Submitted}.
This training round does not significantly affect the present work, and thus
is ignored for the sake of simplicity.}  Imperfect CSI results in non-zero
interference coefficients $|\hv_k^H \hat{\vv}_j|$, which in turn decrease the rate.
In \cite{Submitted} it is shown that the rate in (\ref{eq-rate}) is accurately
lower-bounded by
\begin{eqnarray} \label{eq-rate2}
R^{\rm ZF} - \overline{\Delta R}
\end{eqnarray}
where $R^{\rm ZF}$ is the rate achievable with perfect CSI and $\overline{\Delta R}$ denotes the rate gap given by
\begin{eqnarray}
\overline{\Delta R} \triangleq \log \left(1 + \frac{\rho}{\nt} \sum_{j \ne k}
\mathbb{E} \left[  |\hv_k^H \hat{\vv}_j|^2 \right] \right).
\end{eqnarray}
The rate gap depends on $T_1$, $\tfb$ and the feedback strategy. Its closed-form expressions are found in \cite{Submitted} for the cases addressed in this
paper.

%%%%%%%%%%%%%%%%%%%%%%%%%%%%%%%%%%%%%%%%%%%%%%%%%%%%%%%%%%
\section{Optimizing training and feedback}\label{sect:Optimization}

We now consider the problem of interest, which is the maximization of the \textit{net} spectral efficiency:
\begin{equation} \label{eq-mainopt}
\max_{T_1, \tfb: T_1 + \tfb \leq T}     \left(1-\frac{T_1 + \tfb}{T}\right) \left( R^{\rm ZF} - \overline{\Delta R}(T_1, \tfb) \right).
\end{equation}
To facilitate solving this optimization, it is useful to write our problem as follows:
\begin{equation} \label{eq-mainopt2step}
\max_{T_t \leq T} ~ \max_{T_1 + \tfb = T_t}   \left(1-\frac{T_1 + \tfb}{T}\right) \left( R^{\rm ZF} - \overline{\Delta R}(T_1, \tfb) \right).
\end{equation}
Furthermore, we write the rate gap as follows:
\begin{equation}
 \overline{\Delta R}(T_1, \tfb) = \log \left(1 + g(T_1, \tfb) \right)
\end{equation}
where the function $g(\cdot)$ depends on the feedback strategy and is defined later.
Because the first multiplicative term is constant when $T_1 + \tfb = T_t$ , the inner maximization corresponds to
minimization of the function $g(\cdot)$ subject to the constraint $T_1 + \tfb \leq T_t$:
\begin{equation} \label{eq-opt1}
g(T_t) \triangleq \min_{T_1 + \tfb \leq T_t} g(T_1, \tfb),
\end{equation}
while the second step is a maximization of the net spectral efficiency over $T_t$ (the total training and feedback
symbols):
\begin{equation} \label{eq-mainopt2}
\max_{T_t: ~ T_t \leq T}     \left(1-\frac{T_t}{T}\right) \left( R^{\rm ZF} - \log(1 + g(T_t) \right).
\end{equation}
In the following this two-step strategy is implemented for analog feedback, TDD systems with
channel reciprocity, and digital feedback (with and without feedback channel errors).

%%%%%%%%%%%%%%%%%%%%%%%%%%%%%%%%%%%%%%%%%%%%%%%%%%%%%%%%%%%%%%%%%%%%%%%%%%%%%%%%%%%%%%%%%%%%%%%%%%%%%%%%%%%%%%%%%%%%%%5
\subsection{Analog Feedback} \label{sec:analog}

We begin by considering unquantized analog feedback, whereby the complex amplitude of each discrete-time
feedback symbol is chosen as the UT's estimate of each complex channel coefficient.  Because
each UT is allowed $\frac{\tfb}{\nt}$ feedback channel uses, this corresponds to
$\frac{\tfb}{\nt^2}$ feedback channel uses per channel coefficient (if $\tfb > \nt^2$, each
coefficient is effectively repeated $\frac{\tfb}{\nt^2}$ times on the feedback channel).
This results in distortion that is inversely proportional to $\rho \frac{\tfb}{\nt^2}$, and
the resulting rate gap is described as \cite[Section IV]{Submitted}:
\begin{equation}
g^{\rm analog}(T_1, \tfb) = \frac{\nt - 1}{T_1} + \frac{\nt(\nt - 1)}{\tfb}.
\end{equation}

We begin by minimizing $g(\cdot)$ subject to a constraint on $T_1 +
\tfb$.  For the sake of generality, we rewrite $g()$ as:
\begin{equation}
g^{\rm analog}(T_1, \tfb) = \frac{w_1}{T_1} + \frac{w_{\rm fb}}{\tfb}.
\end{equation}
where $w_1 = \nt - 1$ and $w_{\rm fb} = \nt(\nt-1)$.
Therefore, the minimization to be solved is:
\begin{eqnarray}
\min & \frac{w_1}{T_1} + \frac{w_2}{\tfb} \\
\textrm{subject to} & T_1 + \tfb \leq T_t.
% & T_1 \geq T_1^{\rm min}, ~~ \tfb \geq T_2^{\rm min}.
\end{eqnarray}
%where $T_1^{\rm min} = \nt$ and $T_2^{\rm min} = \nt^2$.
This is readily seen to be a convex optimization, and can be
solved by forming the Lagrangian:
% with respect to the constraint $T_1 + \tfb \leq T_t $:
\[ \Lc(T_1,\tfb,\mu) = g(T_1, \tfb)+\frac{1}{\mu^2} (T_1 + \tfb) \]
where $\mu>0$ is the Lagrangian multiplier. The KKT condition yields
the following solution
\begin{equation}\label{waterfilling}
T_1^{\star} = \sqrt{w_1} \mu, ~~~ \tfb^{\star} = \sqrt{w_{\rm fb}} \mu.
%T_i^{\star} = \max\left\{T_i^{\min}, \sqrt{w_i} \mu \right\}
\end{equation}
In terms of $T_t$, these can be written as
\begin{equation}\label{OptimalT}
 T_1^{\star}  = \sqrt{\frac{w_1}{\Kc}} T_t, ~~~ T_{\rm fb}^{\star}  = \sqrt{\frac{w_{\rm fb}}{\Kc}} T_t
\end{equation}
where we let $\Kc = (\sqrt{w_1} + \sqrt{w_{\rm fb}})^2 $, while the objective value is given by
\begin{equation}\label{OptimalG}
 g(T_t)=  \frac{\Kc}{T_t}
\end{equation}
It is clear that $T_t$ is shared between training
and feedback proportional to the square root of the weights $w_1$ and $w_{\rm fb}$.

Using (\ref{OptimalG}), the overall optimization can now
be characterized in terms of a single variable $T_t$. Namely the
second step of the proposed optimization corresponds to maximizing
\begin{eqnarray}\label{OptimizeTr}
f(T_t)=\left(1-\frac{T_t}{T}\right) \left[R^{\rm ZF}- \log\left(1+ \frac{\Kc}{T_t}\right)\right]
\end{eqnarray}
Because $f$ is concave in $T_t$,  the optimal $T_t^{\star}$
can be found by numerically solving for $\frac{\partial f}{\partial T_t}=0$ where
\begin{eqnarray}\label{Gradient}
\frac{\partial f}{\partial T_t} =
\frac{\Kc\left(1-\frac{T_t}{T} \right)}{T_t^2 \left(1+ \frac{\Kc}{T_t}\right)}
-\frac{1}{T} \left[R^{\rm ZF} - \log\left(1+ \frac{\Kc}{T_t}\right)\right].
\end{eqnarray}

Although a closed-form solution for $T_t^\star$ does not exist, it is possible to
compute how this quantity scales with blocklength $T$.
From (\ref{Gradient}), the optimal $T_t^{\star}$ satisfies the following equality
\begin{equation}\label{kkt}
\frac{\Kc(T-T_t)}{T_t^2\left(1+ \frac{\Kc}{T_t}\right)} =
R_k^{\rm ZF}- \log\left(1+ \frac{\Kc}{T_t}\right)
\end{equation}
It is easy to see that the derivative in (\ref{Gradient}) is upperbounded  by $\frac{1}{T}
\widetilde{f}(T_t)$, where
\begin{equation} \label{gradient-upperb}
\widetilde{f}(T_t) =
\frac{\Kc\left(T - T_t\right)}{T_t^2} - \left[ R^{\rm ZF} - \frac{\Kc}{T_t} \right]
\end{equation}
Since $f$ is concave, it follows that the solution $\widetilde{T}_t$ of the equation $\widetilde{f}(T_t) = 0$ is an upper bound to the optimal value
$T^\star_t$.
Solving $\widetilde{f}(T_t) = 0$ we find
\begin{eqnarray}\label{TrScale}
T_t^{\star}  \leq \widetilde{T}_t = \sqrt{\frac{\Kc T}{R^{\rm ZF}}}
\end{eqnarray}
Furthermore, when the rate gap is small such that $\log\left(1+ \frac{\Kc}{T_t}\right) \approx \frac{\Kc}{T_t}$
(which becomes accurate for large $T$), the upperbound also becomes a very good approximation.

The upperbound (\ref{TrScale}) yields two interesting behaviors: 1)
for a fixed SNR (i.e.,  constant $R^{\rm ZF}$) $T^{\star}_t$
increases as $O(\sqrt{T})$ as $T \rightarrow \infty$;  2) for a
fixed coherence interval $T$, $T^{\star}_t$ decreases as
$O(1/\sqrt{R^{\rm ZF}})$ for large SNR, or equivalently, it decreases
as $O(1/\sqrt{\log(\SNR)})$ since $R^{\rm ZF} =
\log(\SNR)+O(1)$ for large SNR.

In addition, an upper bound on $T_1^{\star}$ can be reached by combining (\ref{TrScale})
with (\ref{OptimalT}):
\begin{eqnarray}\label{T1}
\widetilde{T}_1 = \sqrt{\frac{w_1}{\Kc}} \widetilde{T}_t = \sqrt{\frac{w_1 T}{R^{\rm ZF}}} =
\sqrt{\frac{(\nt - 1) T}{R^{\rm ZF}}}.
\end{eqnarray}
According to this approximation, the optimal downlink training is independent of
$w_{\rm fb}$, and thus of the efficiency of the feedback channel.

Next, we examine the impact of $T_t^{\star}$ on the achievable rate.
Using the upperbound (\ref{TrScale}) into (\ref{OptimizeTr}), the objective value can be approximated as
\begin{eqnarray*}\label{ApproxObj}
f(\widetilde{T}_t) & = & \left(1 - \sqrt{\frac{\Kc}{R^{\rm ZF}
T}}\right) \left[R^{\rm ZF} - \log \LB 1+\sqrt{\frac{ \Kc R^{\rm
ZF}}{T}}\RB \right] \nonumber
\end{eqnarray*}
After some manipulation, it can be shown that the resulting effective rate gap with respect to
$R^{\rm ZF}$ is given by
\begin{eqnarray}\label{EffectiveGap}
R^{\rm ZF} -    f(T_t^{\star}) & \leq & R^{\rm ZF} -    f(\widetilde{T}_t) \approx 2 \sqrt{\frac{\Kc R^{\rm ZF}}{T}}
\end{eqnarray}
Thus, the gap to a perfect CSI system decreases roughly as $O(1/\sqrt{T})$ as $T$
increases.

%%%%%%%%%%%%%%%%%%%%%%%%%%%%%%%%%%%%%%%%%%%%%%%%%%%%%%%%%%%%%%%%%%%%%%%%%%%%%%%%%%%
\subsection{Time-Division Duplexing}

The analysis from the previous subsection can also be used to optimize the amount of uplink training
performed in a time-division-duplexed (TDD) system with perfect channel reciprocity
(i.e., the downlink and uplink channels are identical).\footnote{Note that a similar optimization is
considered in \cite{jose:csi}, although in that work analysis of this optimization is not performed.}
Note that no feedback is necessary in such case.
In \cite[Remark 4.2]{Submitted} the rate gap for a TDD system that uses $T_{\rm TDD}$ uplink training symbols
($\frac{T_{\rm TDD}}{\nt}$ per MS) is given:
\begin{equation}
\overline{\Delta R} = \log \left(1 + \frac{\nt - 1}{T_{\rm TDD}} \right).
\end{equation}
The optimization over $T_{\rm TDD}$ is
\begin{eqnarray}
\max_{T_{\rm TDD} \leq T} \left(1-\frac{T_{\rm TDD}}{T}\right) \left[R^{\rm ZF}- \log\left(1+ \frac{\nt - 1}{T_{\rm TDD}}\right)\right],
\end{eqnarray}
which is clearly equivalent to the optimization in (\ref{OptimizeTr})
with $T_t =T_{\rm TDD}$ and $\Kc = \nt - 1$.  As a result, the analysis and approximations from the previous subsection carry over.
By adapting (\ref{TrScale}) we have
\begin{eqnarray}
T_{\rm TDD}^{\star}  \leq \widetilde{T}_{\rm TDD} = \sqrt{\frac{(\nt - 1) T}{R^{\rm ZF}}},
\end{eqnarray}
which is the same as the approximation to $T_1^\star$ (the number of downlink training symbols) for
analog feedback in (\ref{T1}).

Based upon the expression for the approximate rate gap in (\ref{EffectiveGap}), by comparing the value of $\Kc$
for analog feedback and
for TDD we see that the rate gap for analog feedback is a factor $1 + \sqrt{\nt}$ larger than for TDD.

For future reference it is also worthwhile to notice that the TDD setting corresponds to the non-TDD setting with perfect feedback (i.e., the BS knows the
UT channel estimates, or equivalently $w_{\rm FB}=0$ in Section \ref{sec:analog}). As a result, the net rate achievable with TDD serves as an upper bound
to that achievable with training and channel feedback.

%%%%%%%%%%%%%%%%%%%%%%%%%%%%%%%%%%%%%%%%%%%%%%%%%%%%%%%%%%%%%%%%%%%%%%%%%%%%%%%%%%%%%%%%%%%%5
\subsection{Error-Free Digital Feedback}

We now analyze digital feedback techniques, whereby each UT quantizes its vector channel estimate to
$B$ bits and then maps these bits into $\frac{\tfb}{\nt}$ transmit symbols. For the quantization step we
consider a family of random vector quantization (RVQ) schemes. Assuming the feedback bits are received
error-free, in \cite[Section V]{Submitted} it is shown that the rate gap is
\begin{equation} \label{deltaR_digital}
\overline{\Delta R} = \log \left(1 + \frac{\nt - 1}{T_1} + \rho ~ 2^{-\frac{B}{\nt - 1}}  \right)
\end{equation}
%This expression is in terms of bits, whereas we require an expression in terms of feedback symbols.
where the distortion error is expressed in terms of bits.
In this section we assume unrealistically that error-free communication
is possible over the feedback channel at its underlying capacity of $\log_2 \left(1 + \frac{P}{N_0} \right)$ bits
per channel use.  Each of the $\nt$ UT's utilize $\frac{\tfb}{\nt}$ channel uses, and therefore
$B = \frac{\tfb}{\nt}\log_2 \left(1 + \rho \right)$.  As a result, we obtain
\begin{equation} \label{g-digital}
g^{\rm digital}(T_1, \tfb) = \frac{\nt - 1}{T_1} + \rho \left(1 + \rho \right)^{- \frac{ \tfb }{\nt (\nt - 1)}}.
\end{equation}
The first step is the minimization of the above function subject to the constraint
$T_1 + \tfb \leq T_t$. Since $g^{\rm digital}$ is convex in $T_1, \tfb$, we form the Lagrangian and
readily obtain
\begin{eqnarray} \label{Tfb-errorfree}
T_1 &=& \mu \sqrt{\nt - 1} \\
\tfb &=& \nt(\nt-1) \frac{ 2 \ln(\mu) + \ln \left(\frac{ \rho \ln(1+\rho) }{\nt (\nt-1)}\right)}{\ln (1 + \rho)}
\end{eqnarray}
Note that the feedback length grows as $O\left(\ln \mu\right)$, much slower than the linear increase
(in $\mu$) for the common training.

Contrary to the earlier analog feedback case, $g^{\rm digital}$ cannot be expressed as a closed form of $T_t$ but instead must be expressed as a function
of $\mu$.  However, for the sake of comparison with analog feedback
we perform this optimization in terms of $T_1$ rather than $\mu$. Based upon (\ref{Tfb-errorfree}) we can express $\tfb$ as a function of $T_1$:
\begin{eqnarray}
\tfb = \nt(\nt-1) \frac{ 2 \ln(T_1) + \ln \left(\frac{ \rho \ln(1+\rho) }{\nt (\nt-1)^2}\right)}{\ln (1 + \rho)},
\end{eqnarray}
and thus the net spectral efficiency can be written as:

\begin{eqnarray} \nonumber
%    f^{\rm digital} (T_1) &=& \left(1-\frac{T_t^{\rm DF-awgn}(\mu)}{T}\right) \times \\
\left(1-\frac{T_1 +  \nt(\nt-1) \frac{ 2 \ln(T_1) + \ln \left(\frac{ \rho \ln(1+\rho) }{\nt (\nt-1)^2}\right)}{\ln (1 + \rho)}}{T}\right) \times \\
    \left[R^{\rm ZF}- \log\left(1+ \frac{\nt - 1}{T_1} + \frac {\nt(\nt - 1)^2}{(T_1)^2 \ln(1 + \rho)} \right) \right].
\end{eqnarray}
Because $\tfb$ increases logarithmically with $T_1$, its effect on the maximization is rather minimal.  As a result, the
maximization of $T_1$ is very similar to the maximization of $T_{\rm TDD}$ in the TDD setting, which is in turn similar to the maximization of $T_1$ in
the presence of analog feedback.

%%%%%%%%%%%%%%%%%%%%%%%%%%%%%%%%%%%%%%%%%%%%%%%%%%%%%%%%%%%%%%%%%%%%%%%%%%%%%%%%%%%%%%%%%%%%5
\subsection{Digital Feedback with Errors}

Rather than assuming (unrealistically) that the feedback channel operates at channel capacity and
error-free, in this section we analyze a system where uncoded QAM is used to transmit each UT's quantized channel vector
over the feedback channel. Each UT utilizes $\frac{\tfb}{\nt}$ feedback channel uses.  Assuming that quantization bits
are arbitrarily mapped to channel symbols, one or more symbol errors (among the $\frac{\tfb}{\nt}$ channel uses)
makes the
feedback from a particular UT effectively useless and thus leads to a rate effectively of zero.
Under this assumption, the achievable net rate is given by:
\begin{eqnarray}\label{errorDF-objective}
   \left(1-\frac{T_1 + \tfb}{T}\right) (1-\overline{P}_{e,\rm fb})
    \left[R_k^{\rm ZF} - \overline{\Delta R} \right]
\end{eqnarray}
where $\overline{\Delta R}$ is defined in (\ref{deltaR_digital}).  Because each UT utilizes $\frac{\tfb}{\nt}$ complex channel symbols,
the number of feedback bits per user $B = \frac{\tfb}{\nt} \log_2 M$ where $M$ is the number of constellation points. The
per-symbol QAM error probability is given by
\begin{equation}
P_s = 1 - \left(1 - 2 \left(1 - \frac{1}{\sqrt{M}}\right) Q \left( \frac{3 ( P/N_0)}{M-1} \right) \right)^2,
\end{equation}
while the probability of a feedback error, $\overline{P}_{e,\rm fb}$, is the probability that any of the symbols are received incorrectly:
\begin{equation}
\overline{P}_{e,\rm fb}  = 1 - (1 - P_s)^{\frac{\tfb}{\nt}}.
\end{equation}

In order to allow for a two-step optimization, we rewrite the objective in (\ref{errorDF-objective}) as:
\begin{eqnarray}\label{errorDF-objective2}
   \left(1-\frac{T_t}{T}\right)  \left[R_k^{\rm ZF} - h(T_t) \right]
\end{eqnarray}
where the effective rate-loss $h(T_t)$ incorporates the loss due to feedback error is defined as:
\begin{eqnarray}
h(T_t) = \min_{T_1, \tfb: T_1 + \tfb \leq T_t} w(T_1, \tfb)
\end{eqnarray}
with
\begin{eqnarray} \nonumber
w(T_1, \tfb) \! \!\!\!\! &=& \!\!\!\!\! \left(1 - \overline{P}_{e,\rm fb} \right)
\log \left(1 + \frac{\nt - 1}{T_1} + \rho M^{-\frac{\tfb}{\nt(\nt - 1})}  \right) \\ &&
+ \overline{P}_{e,\rm fb} R_k^{\rm ZF}.
\end{eqnarray}
If a reasonable constellation size is used, the probability of feedback error is quite small
even when the number of feedback bits per users is relatively large
(e.g., for $\nt=4$ at $10$ dB with $B=25$ and 4-QAM, $\overline{P}_{e,\rm fb} = 0.038$).
As a result, the minimization of $w(T_1, \tfb)$ is very similar to the minimization of
$g^{\rm digital}(T_1, \tfb)$ for error-free feedback in (\ref{g-digital}), but with a
constellation of size $M$ rather than $1 + \rho$.  When computed numerically, an optimization
over the constellation size is also performed.

%%%%%%%%%%%%%%%%%%%%%%%%%%%%%%%%%%%%%%%%%%%%%%%%%%%%%%%%%%
\section{Numerical Examples \& Discussion}

This section provides some numerical examples to illustrate the
analysis of the previous section.  The system parameters, unless
otherwise specified, are $\nt = 4$ and $\rho = 10$ dB. In all cases
the previously stated optimizations have been numerically computed with
$T_1$ and $\tfb$ restricted to integers, subject to the constraints
$T_1 \geq \nt$ (to ensure at least one training symbol per channel coefficient) and $\tfb \geq \nt^2$ for
analog feedback (one feedback symbol per channel coefficient) and $\tfb \geq \nt$ for digital
feedback (one FB symbol per UT).

\begin{figure}
    \begin{center}
   \epsfxsize=3.0in
   \epsffile{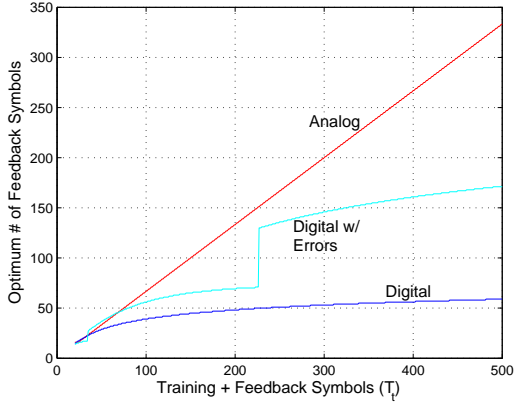}
    \end{center}
    \caption{Optimum number of feedback symbols ($\tfb$) versus total training + feedback ($T_t$).}
    \label{fig:opt_tfb}
\end{figure}

In Fig. \ref{fig:opt_tfb} the optimum number of feedback symbols ($\tfb$) is plotted versus $T_t$, the total training \& feedback budget, for analog
feedback, digital feedback, and digital feedback with errors (uncoded QAM).  For analog feedback the number of feedback symbols grows linearly with $T_t$
with slope $\tfb = \frac{\sqrt{\nt}}{1 + \sqrt{\nt}}$, while for digital feedback $\tfb$ increases with $T_t$ at a much slower rate (approximately
logarithmically). The rate of increase for $\tfb$ is particularly slow beyond $T_t = 100$.  At this point digital feedback without errors corresponds to
$10$ symbols per user and thus nearly $35$ bits ($B$).  At this point the distortion due to quantization ($2^{-\frac{B}{\nt - 1}}$) is less than $10^{-3}$
and the gains in increasing $B$ beyond this point are very negligible. Even when feedback using uncoded 4-QAM is considered, each user is quantizing to
$28$ bits at $T_t = 100$. The abrupt shift for digital feedback with errors occurs when the constellation changes from 4-QAM to BPSK: when the number of
feedback symbols becomes too large (when $T_t$ is sufficiently large) the probability of feedback error becomes significant and it becomes more efficient
to reduce this error probability by reducing the constellation to BPSK while keeping the number of bits ($B$) nearly the same.  This is a consequence of
using uncoded transmission on the FB channel.

\begin{figure}
    \begin{center}
   \epsfxsize=3.0in
   \epsffile{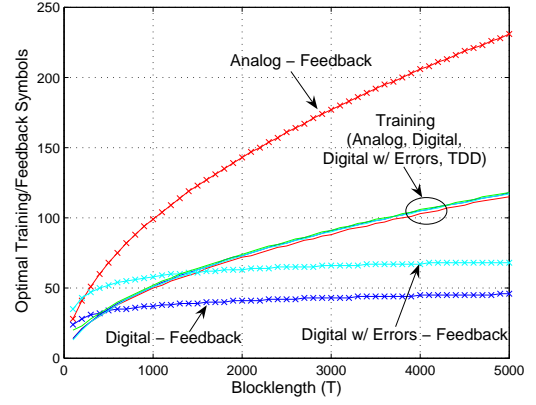}
    \end{center}
    \caption{Optimum number of feedback symbols ($\tfb$) and training symbols ($T_1$) versus blocklength ($T$).}
    \label{fig:opt_fb_train}
\end{figure}

In Fig. \ref{fig:opt_fb_train} the optimal values of $T_1$ and $\tfb$ are plotted versus
blocklength $T$ for analog, digital, and digital w/ errors; $T_{\rm TDD}$ is also plotted for TDD.
Most striking is the fact that the optimal values of $T_1$ and the optimal $T_{\rm TDD}$
are essentially identical for the three feedback techniques as well as for TDD.
Furthermore, although not shown here, the optimizing values of $T_1$ are
very well approximated by $\sqrt{\frac{(\nt - 1) T}{R^{\rm ZF}}}$ as in (\ref{T1}).
On the other hand, the number of feedback symbols depends critically on the feedback method.
Because analog feedback is so inefficient, a large number of
feedback symbols are used so that the rate gap due to feedback is not too large.
On the other hand, digital feedback is very efficient and a relatively small number of
feedback symbols is required.

\begin{figure}
    \begin{center}
   \epsfxsize=3.0in
   \epsffile{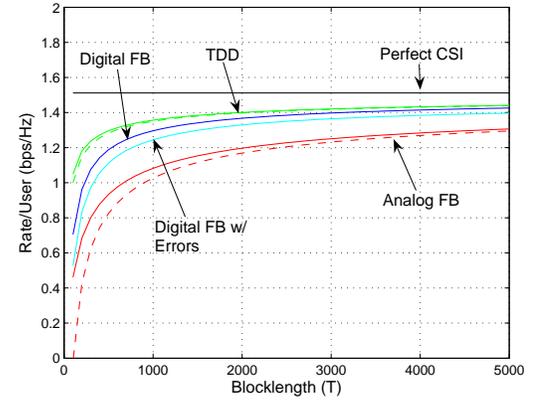}
    \end{center}
    \caption{Achievable sum rate vs. blocklength ($T$).}
    \label{fig:rate_vs_T}
\end{figure}

In Fig. \ref{fig:rate_vs_T} the net achievable rate is plotted versus blocklength $T$.  For analog
and TDD the rate approximations based upon (\ref{EffectiveGap}) are indicated with dotted lines and
are seen to become increasingly accurate as $T$ is increased.
Analog feedback is outperformed
by digital feedback, with or without errors, for all blocklengths.  This is because
digital feedback offers a significantly smaller distortion as compared to analog
whenever $\tfb$ is larger than (approximately) $\nt^2$ (i.e., one symbol per channel coefficient)
\cite[Section VI]{Submitted}, and for reasonable blocklengths it is optimal to use
$\tfb$  larger than $\nt^2$ (Fig. \ref{fig:opt_fb_train}).

%%%%%%%%%%%%%%%%%%%%%%%%%%%%%%%%%%%%%%%%%%%%%%%%%%%%%%%%%%

%%%%%%%%%%%%%%%%%%%%%%%%%%%%%%%%%%%%%%%%%%%%%%%%%%%%%%%%%%
%\bibliographystyle{IEEEtran}
%\bibliography{asilomar06}

\end{document}